\documentclass[11pt]{report}
\usepackage{amsfonts}
\usepackage[final]{pdfpages}
\usepackage{ucs}
\usepackage{color,graphicx}
\usepackage{amssymb}
\usepackage{wrapfig}
\usepackage{setspace}
\usepackage{version}
\usepackage{fullpage}
\usepackage{framed}
\usepackage[color]{changebar}
\usepackage{program}
\usepackage[authoryear]{natbib}
\usepackage{graphicx}
\usepackage{caption}
\usepackage{subcaption}

\cbcolor{blue}

\begin{document}

\begin{flushleft}
{\Large \textbf{A Markov chain model of evolution in asexually reproducing populations: insight and analytical tractability in the evolutionary process}}

Daniel Nichol$^1$, Peter Jeavons$^1$, Robert Bonomo$^2$, Philip K. Maini$^3$, Jerome L. Paul$^4$, Robert A. Gatenby$^5$, Alexander R.A.\ Anderson$^5$ \& Jacob G.\ Scott$^{3,5}$\\
\vspace*{0.5cm}
\small{
$^1$ Department of Computer Science, University of Oxford, Oxford, UK\\
$^2$ Department of Medicine, Louis Stokes Department of Veterans Affairs Hospital, Cleveland, OH, USA \\
$^3$ Centre for Mathematical Biology, University of Oxford, Oxford, UK\\
$^4$ School of Computing Sciences and Informatics, University of Cincinnati, Cincinnati, OH, USA\\
$^5$ Integrated Mathematical Oncology, H. Lee Moffitt Cancer Center and Research Institute, Tampa, FL, USA\\
}
{\normalsize \vfill
Correspondence:\newline
Jacob G Scott, e-mail: jacob.g.scott@gmail.com and \newline
Daniel Nichol, e-mail: daniel.nichol@st-annes.ox.ac.uk }

{\normalsize 
}
\end{flushleft}

\newpage

\section*{Abstract}

The evolutionary process has been modelled in many ways using both stochastic and deterministic models.  We develop an algebraic model of evolution in a population of asexually reproducing organisms in which we represent a stochastic walk in phenotype space, constrained to the edges of an underlying graph representing the genotype, with a time-homogeneous Markov Chain.  We show its equivalence to a more standard, explicit stochastic model and show the algebraic model's superiority in computational efficiency.  Because of this increase in efficiency, we offer the ability to simulate the evolution of much larger populations in more realistic genotype spaces.  Further, we show how the algebraic properties of the Markov Chain model can give insight into the evolutionary process and allow for analysis using familiar linear algebraic methods.


\newpage

\section*{Introduction}

Understanding the evolutionary trajectories of populations under given selective pressures is a fundamental problem in biology. In particular, understanding which traits are likely to be selected for within a population can help us better treat disease and understand ecological changes. Traditional computational models of evolution have often simulated populations explicitly (\cite{moran1962statistical}) - keeping a population of members and probabilistically iterating phases of reproduction, mutation and selection. These methods are cumbersome and can become too computationally complex to simulate for large populations or large numbers of accessible genotypes. 

Past successful simplifications of this evolutionary process for asexually reproducing organisms have been to model a genotypically homogeneous population which undertakes a stochastic walk of mutations, mutating at each time step to a fitter variant as a population. This model makes the so-called Strong Selection Weak Mutation assumption and has been used to study which evolutionary trajectories are inaccessible to a population of organisms and how these trajectories are changed by sign epistasis - the situation in which a given mutation may be beneficial under certain selective pressures but deleterious under others (\cite{weinreich2005perspective, Weinreich:2006uq, tan2011hidden, poelwijk2011reciprocal}). Certainly this model is efficient to simulate, but it tells us little about population dynamics. In particular, how do different selective pressures influence how a heterogeneous population diversifies or converges over time?

In this paper we present two models of evolution: one an explicit stochastic model which tracks each individual within the population and simulates mutation and selection, and the other a more abstract description using a time-homogeneous Markov chain.  We show that the two models, given the same evolutionary landscape and initial population, result in nearly identical populations.  Further, once the evolutionary process has been encoded in a Markov chain, we show that there is additional insight which can be gained by familiar linear algebraic analysis.  Finally, we show that use of this novel method of encoding the evolutionary process is far less computationally intensive than more standard explicit models and hence allow for exact calculation of evolution for large populations on large, rugged evolutionary landscapes.


\section*{Fitness Landscapes and the Genotype-Phenotype Map}

The mapping from genotype to phenotype is a familiar concept, but one that has eluded rigorous treatment in the genomic era.  While mutation certainly occurs at the level of the genotype, selection operates at the level of the phenotype, making this mapping central to any study of evolution.  We will begin the process of modelling this mapping by utilizing the representation of the genotypes of an asexually reproducing organism as presented by Weinreich and colleagues to study evolutionary trajectories (\cite{weinreich2005perspective}).  Once we have established a framework for the genotype, which constrains the allowable paths through mutation space, we will invoke a genotype-phenotype mapping to establish the `forces' of selection.  This two-level system will then constitute the basis for our models. 

\begin{figure}
\centering
\begin{subfigure}[b]{0.48\textwidth}
\centering
\includegraphics[width=\textwidth]{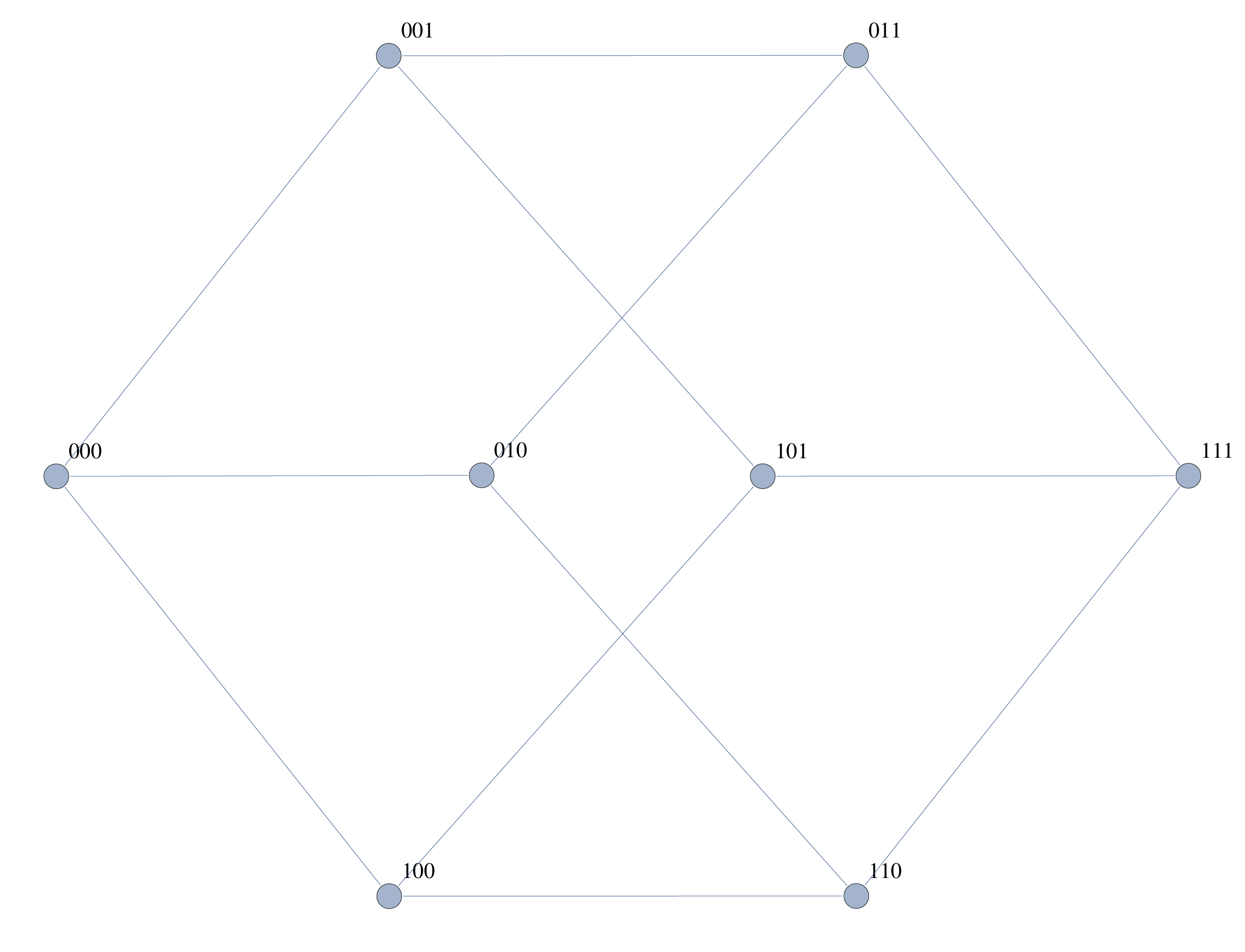}
\caption{} 
\label{fig:gspace}
\end{subfigure}
\begin{subfigure}[b]{0.48\textwidth}
\centering
\includegraphics[width=\textwidth]{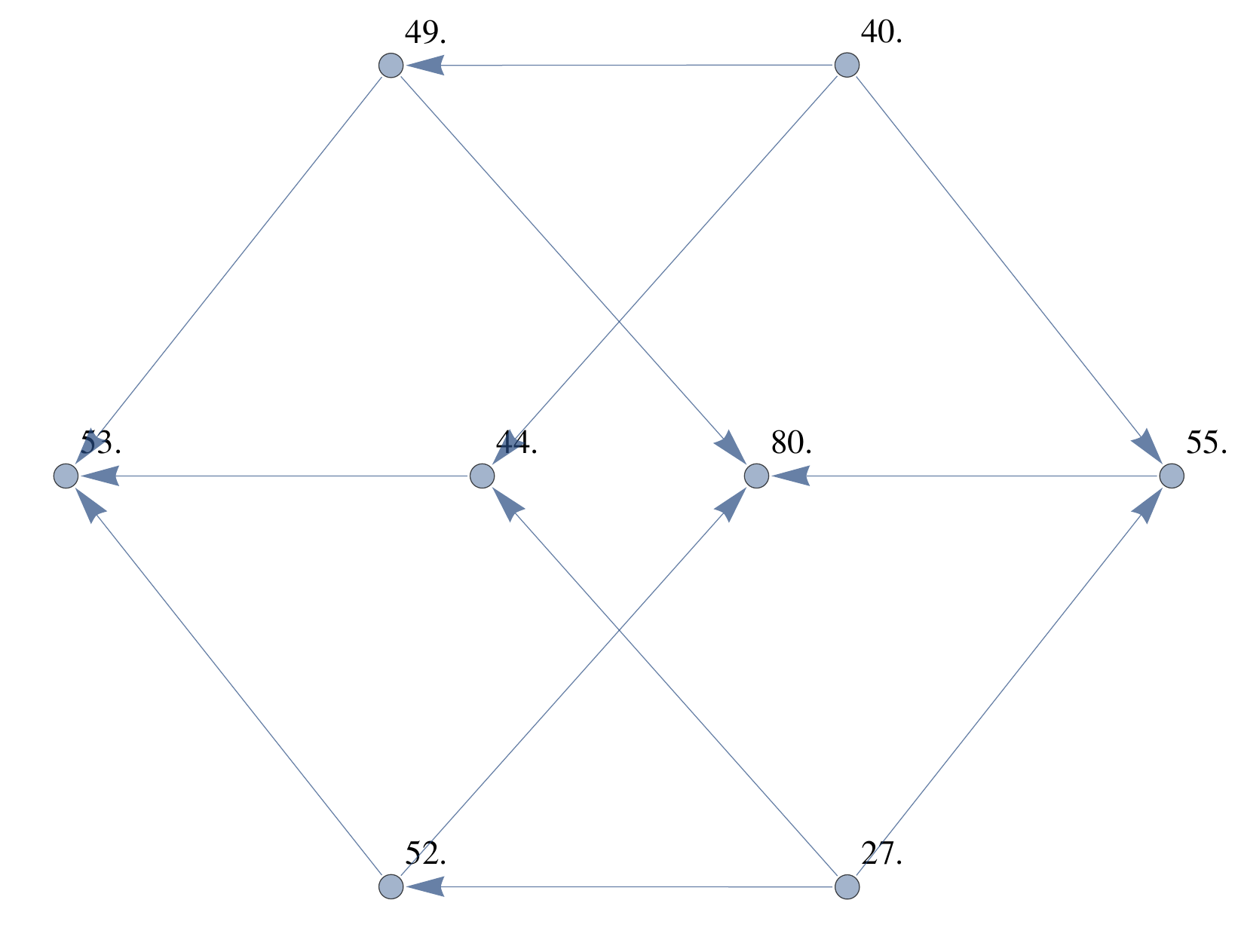}
\caption{} 
\label{fig:pspace}
\end{subfigure}
\caption{(a) The genotype space, $\mathcal{G}$, of an $N=3$ cell, where the genotypes are represented by bit strings of length $N$ and the edges of the graph represent Hamming distance $1$ connections.  (b) An example phenotype graph, $\mathcal{P}$, with vertices represented by integer fitness values determined by the mapping, $f$, from equation \ref{eq:fitness}, and directional edges representing of evolutionarily allowable mutational transitions.}\label{fig:spaces}
\end{figure}

To this end, we represent the genotype of an organism by a bit string of length $N$ and model mutation as the process of flipping a single bit within this string.  This gives a set of possible genotypes of size $2^{N}$ in which each genotype, $x$, has $N$ one-mutation neighbours - precisely those genotypes for which the Hamming distance (\cite{hamming:1950}) from $x$ is $1$. As such, our genotype space ($\mathcal{G}$), can be represented by an undirected $N$-cube graph with vertices which represent genotypes, and edges which connect neighbours at Hamming distance 1 (See Fig. \ref{fig:gspace}).

We then define a selective pressure on our graph that drives evolution, for example through an environmental change or drug application, as a fitness function which acts on each vertex in the graph, $\mathcal{G}$,

\begin{equation}
f:\{0,1\}^{N}\rightarrow \mathbb{R}^{\geq 0}. \label{eq:fitness}
\end{equation}

This fitness function represents a genotype-phenotype map in the simplest sense - assigning to each genotype a single real-valued fitness. This could be, for example, thought of as resistance to an applied drug, as it was by Weinreich (\cite{Weinreich:2006uq}) and later Tan (\cite{tan2011hidden}), to study evolutionary trajectories of \textit{E. Coli} under selection by different beta-lactam antibiotics.

Using these fitness values we construct a directed evolutionary graph on the set of $2^N$ possible genotypes where there exists an edge from $a$ to a neighbour $b$ if and only if $f(b) > f(a)$ (See Fig. \ref{fig:pspace}). This graph provides a model of which mutations and which evolutionary trajectories (series of mutations) are possible - those which increase the fitness of the individual at each step.


\section*{An Explicit Computational Model of Evolution}

\begin{figure}
\includegraphics[width=\textwidth]{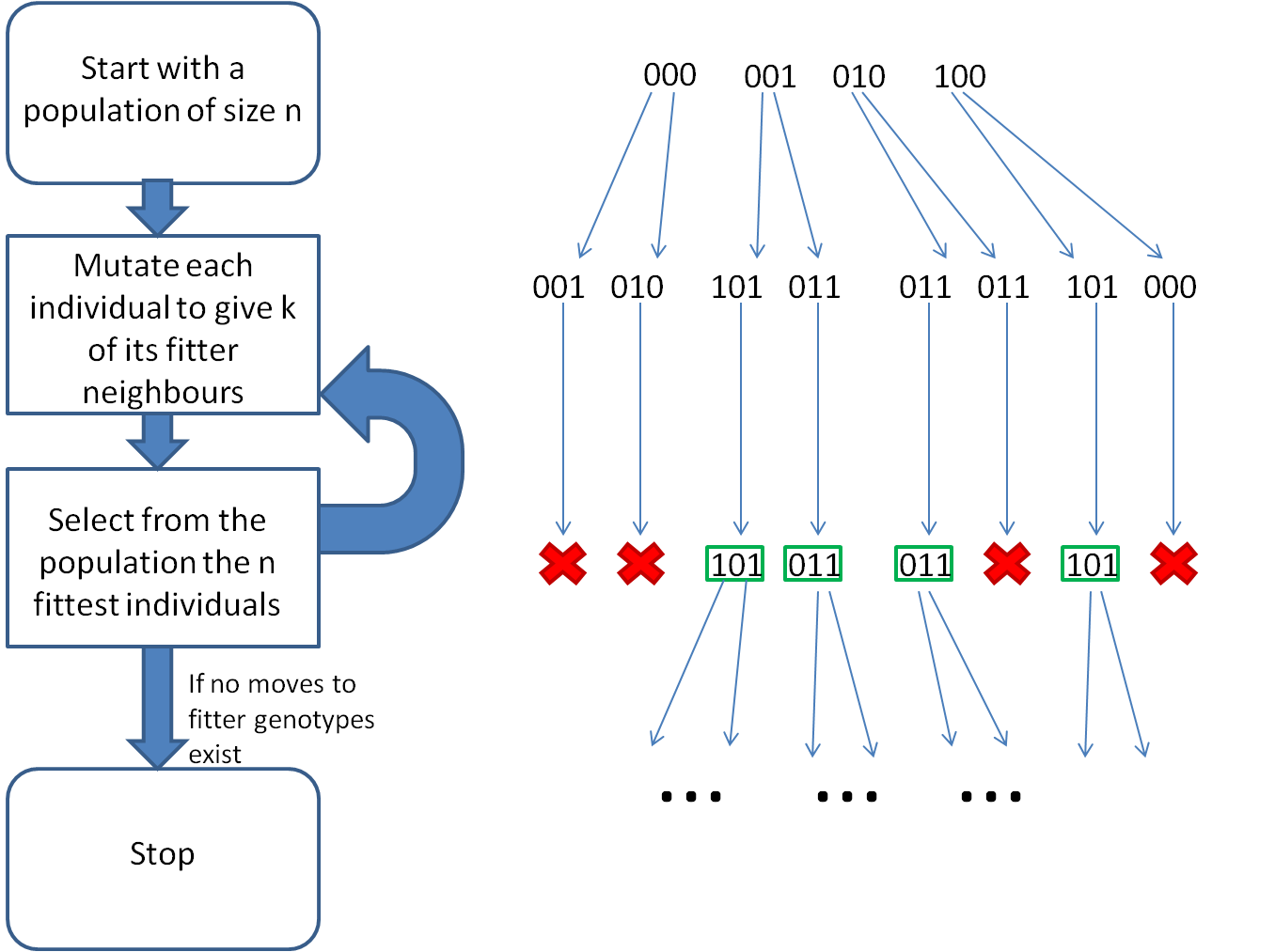}
\caption{An explicit stochastic evolutionary process.  Each individual in a population is mutated at a baseline mutation rate to each of its $k$ fitter neighbours by a stochastic process biased by the difference in fitness.  Selection (green rectangle) acts on the next generation by eliminating the least fit individuals (red X).  The process repeats itself until there are no fitter neighbours to which to mutate.} \label{fig:stochasticmodel}
\end{figure}

Our first attempt to the evolutionary dynamics of this system was by an explicit simulation of a population undergoing mutation and selection over time, similar to the familiar Moran and Wright-Fischer processes (\citep{moran1962statistical,fisher1930}). We keep a population of $n$ individuals and at each time step replace each individual with $k$ of its fitter neighbours. From the $nk$  individuals we have, we choose the fittest $n$ to survive the selection phase. We repeat this process until each individual in the population has no fitter one-mutation neighbours. Figure \ref{fig:stochasticmodel} shows this process for $n=4$ and $k=2$. This model, whilst certainly simplified, attempts to mimic the process of evolution as closely as possible, so we might expect it to make good predictions about the evolutionary trajectories of real-world organisms.

This algorithm requires, at each iteration $nk$, random samples from a uniform distribution. We know that in a random landcape the average mutational path length to a fitness optima is $\log_2(N-1)$ (\cite{altenberg1997nk}). Hence, the expected number of random samples required is $\mathcal{O}(nk\log_2(N-1))$ well as the extra time associated with the sorting in the selection step, which we choose to ignore as it is considerably quicker than the random sampling.

As the genotype spaces of real-world organisms can be very large, this algorithm can require many iterations for the population to stabilize into a final population and to fully explore a large and multi-peaked fitness landscape we would need to maintain a large population. Further, this algorithm must be run many times to give an expected distribution for which the error is acceptable. For example, the simulation required to produce Figure \ref{fig:comparison} required 1000 iterations on a landscape for which $N=5$. The running time of this algorithm makes it intractable for finding expected distributions in genotype spaces which are large enough to be useful.

\section*{A Markov Model of Evolution}

To overcome the problem of inefficiency in explicit simulations and, in particular, the need to iterate them many times to obtain significant results, we will build an algebraic model of the same evolutionary process. We represent the genotype of an organism as a bit string and the phenotype as a specific mapping of each vertex, as before, and model evolution of an organism within a population as an uphill stochastic walk in phenotype space. A member of the population with a genotype $x\in \{0,1\}^{N}$ is replaced in one time step by a fitter one-mutation neighbour with genotype $y$ with probability proportional to increase in fitness (with regard to $f$) and normalised by the other possible evolutionary steps. That is, if a member has genotype $x$, then the probability that it has a different genotype, y, in the next time step
 
\begin{equation}
\Pr (x\rightarrow y)=\begin{cases} \frac{f(y) - f(x)}{\sum_{ Ham(x,z) = 1}
max\{f(z) - f(x), 0\} } & f(y) > f(x)\\ 0 & \text{otherwise} \end{cases} .
\end{equation}

\noindent If we further define, for each genotype $x$, a probability

\begin{equation}
\Pr (x\rightarrow x)=\begin{cases} 1 & \text{if $x$ has no fitter one-step
neighbours} \\ 0 & \text{otherwise} \end{cases} ,
\end{equation}

\noindent our model becomes a time-homogeneous absorbing Markov chain with a finite state space and transition matrix $P=[p_{ij}]$ where $p_{ij}=\Pr (i\rightarrow j)$ for each $i,j\in \{0,1\}^{N}$.

Using this Markov chain we can explore how a population of a given organism comprising a variety of genotypes evolves over time. To do this we make the assumption that the population size is large and remains constant so that only beneficial mutations will fix in the population.  We define a collection of population row vectors $\mu^{(t)}$ for each $t \in \mathbb{N}$, where $\mu^{(t)}$ is a vector of length $2^N$ in which $\mu^{(t)}_{k}$ is the proportion of the population in the $k$th genotype at time $t$. Given such a population at time $t$ the one-step update can be computed by

\begin{equation}
\mu^{(t+1)} = \mu^{(t)}P.
\end{equation}

\noindent As a result of the associativity of matrix multiplication we can compute the distribution of a population at time $t$ given an initial population $\mu^{(0)}$ by

\begin{equation}
\mu^{(t)} = \mu^{(0)}P^t.
\end{equation}

\begin{figure}
\includegraphics[width=\textwidth]{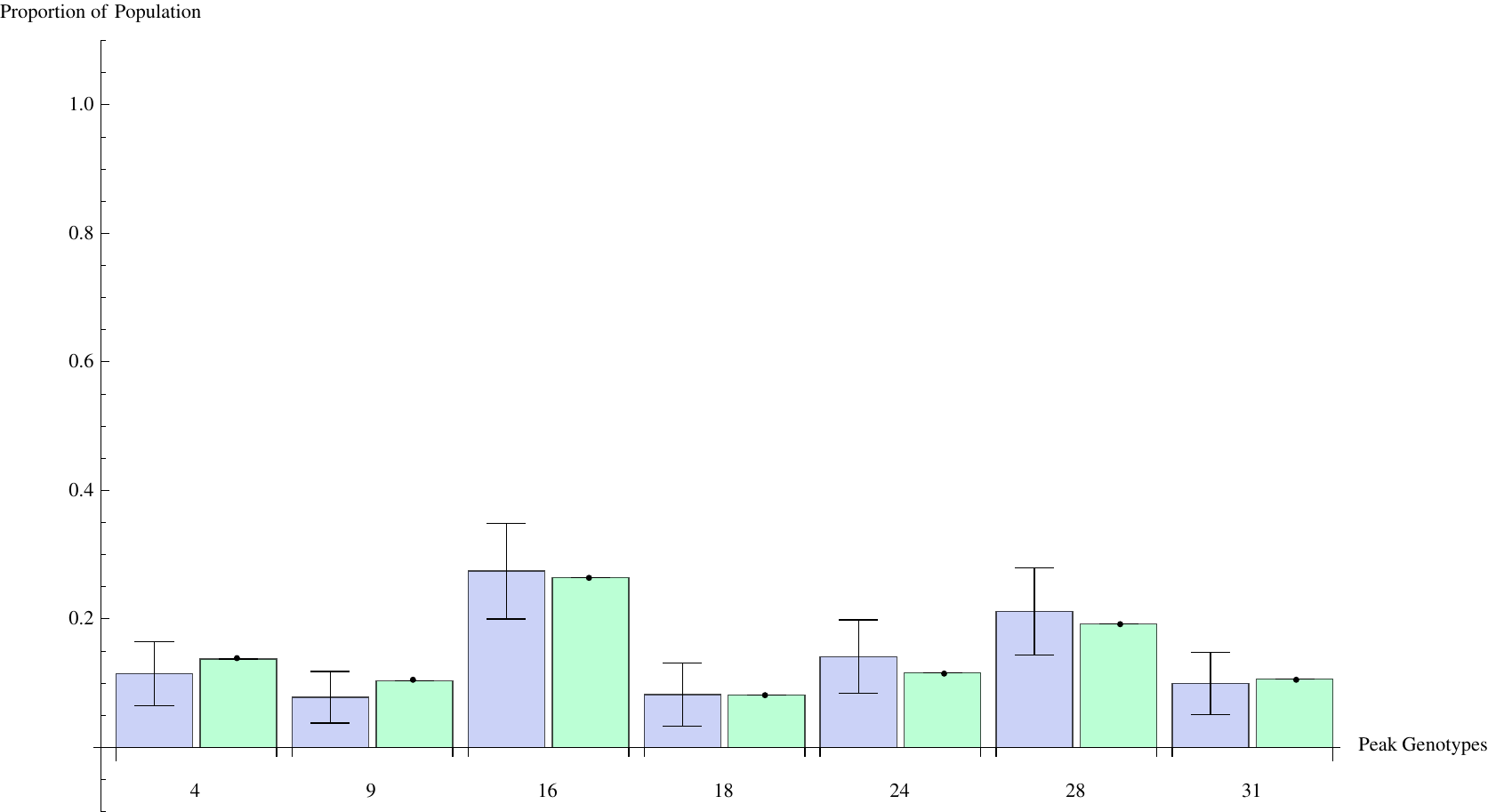}
\caption{A comparison of the two models. For a given randomly generated fitness landscape over a genotype space with $N=5$ we plot the distribution of the population at the 7 possible peaks when each algorithm terminates. The dark blue bars represent the average of 1000 runs of the explicit algorithm with a starting population of size 30. The error bars give the standard error in the predictions of this model. The light blue bars give the population distribution predicted by the Markov chain model}
\label{fig:comparison}
\end{figure}

As we have now encoded the same evolutionary process in two different formalisms, one stochastic and one deterministic, we compare the results of each with identical initial conditions.  Figure \ref{fig:comparison} shows an example output for each model given a single, randomly generated landscape over a genotype space with $N=5$.  The starting populations, upon which the ending populations are highly dependent, are identical.  In this case, the stochastic process was run until termination in all cases, and the deterministic model was calculated to steady state.   We find that the results of the Markov Chain model fall, as expected, within the standard error of the stochastic model.

\section*{Evolution on Large Time Scales}

To explore long term evolutionary properties of populations using an explicit evolutionary algorithm we would have to iterate the population for a large number of generations. For large populations, or for ones with realistic genotype lengths, this can be prohibitively inefficient. In our Markov model of evolution we can simulate each time step much more efficiently and so can explore the long term behaviour much more effectively. While this increase in computational efficiency is a benefit of the model, a more significant improvement is revealed by its algebraic structure. As our evolutionary process is now encoded by a Markov Chain we can explore properties of the process analytically by examining the transition matrix $P$ with no need for simulations at all. The following lemma explores what happens in the evolutionary process over large time scales.

\noindent \textbf{Lemma:}\newline
\emph{\ Let $P$ be a transition matrix as described above. Then there exists
some $k$ such that $P^kP = P^k$ } \newline
\textbf{Proof:}\newline
We first note that, without loss of generality, we may (by rearranging the population vector ordering) assume that our transition matrix $P$ has the block matrix form:

\begin{equation}
P = \left[ 
\begin{array}{cc}
Q & R \\ 
0 & I 
\end{array}
\right],
\end{equation}

\noindent where $Q$ is a transition matrix encoding the probabilites of moving between non-absorbing (i.e. non fitness optima) genotypes, $R$ encodes the probabilites of moving from non-absorbing states into absorbing states, and $I$ is the identity matrix encoding that once in an absorbing state the walk will remain there. Now taking powers of $P$ gives 

\begin{equation}
P^k = \left[ 
\begin{array}{cc}
Q^k & (I + Q + Q^2 + ... + Q^{k-1})R \\ 
0 & I 
\end{array}
\right].
\end{equation}

\noindent Note that if $Q^k = 0$ for some $k$ then $P^kP = P^k$ as

\begin{equation}
P^kP = \left[ 
\begin{array}{cc}
0 & (I + Q + Q^2 + ... + Q^{k-1})R \\ 
0 & I 
\end{array}
\right] \left[ 
\begin{array}{cc}
Q & R \\ 
0 & I 
\end{array}
\right] = \left[ 
\begin{array}{cc}
0 & (I + Q + Q^2 + ... + Q^{k-1})R \\ 
0 & I 
\end{array}
\right] = P^k
\end{equation}

\noindent and so the lemma follows if $Q^{2^N+1} = 0$. Assume to the contrary that $Q^{2^N+1} \ne 0$. Writing $Q^{2^N+1} = [q^*_{i,j}]$ we have for some $n, m$ that $q^*_{n,m} \ne 0$. By the definition of the power of a stochastic matrix the probability of transitioning from the $n$th to $m$th non-absorbing states in $2^N + 1$ steps is non-zero. However, we have by construction that any evolutionary paths amongst those genotypes which are non-absorbing are increasing in fitness at every step and are necessarily acyclic. It follows then, that as $q^*_{n,m} \ne 0$ there exists an acyclic path of length $2^{N}+1$ in a space of fewer than $2^N$ genotypes. This yields a contradiction and the lemma follows. \qed

As a consequence we know that the matrix

\begin{equation}
P^* = \lim_{k\to \infty} P^k
\end{equation}

\noindent exists and in fact this limit is found after only finitely many iterations. It follows that a given initial population $\mu^{(0)}$ will converge to a stationary distribution $\mu^*$ after a finite number of steps in our evolutionary model. Furthermore if $P^*$ is known then we can determine this stationary distribution by the calculation

\begin{equation}
\mu^* = \mu^{(0)}P^*.
\end{equation}

Therefore, the limit matrix $P^*$ need only be calculated once and can be used to explore the behaviour of any number of different starting configurations. This offers an improvement over an explicit evolutionary model as we can compute the limit matrix $P^*$ by repeatedly squaring the transition matrix $P$. The previous lemma shows we need to square $P$ at most $N$ times to find $P^*$. We know that the product of two $m \times m$ matrices can computed in time $\mathcal{O}(m^{2.807})$ by Strassen's algorithm (\cite{strassen1969gaussian}). It follows that this simulation has worst case time complexity $\mathcal{O}(N2^{2.807N})$ in computing $P^*$, although in many cases it is considerably faster. This is slower than a single run of the explicit simulation, but considerably faster than  using the explicit simulation to determine the expected population distribution.

It could be argued that this construction introduces determinism into the inherently stochastic process of evolution but this is not so. The population vector $\mu^*$ represents an expected population distribution after a large number of iterations and gives the same information as running an explicit simulation many times and averaging the results (cf. Fig. \ref{fig:comparison}). 



\section*{Discussion}

We have taken an explicit evolutionary algorithm and encoded it as an equivalent Markov process which reduces each update step of the algorithm to a single matrix multiplication. Further we have seen that we can reduce the problem of determining the evolutionary trajectory of a given starting population to a single matrix multiplication with the matrix $P^*$ which can be efficiently computed.

There is a rich theory of how the form of a matrix determines its properties under multiplication and exponentiation and the previous examples show that these properties can help us gain useful insight into the evolutionary process by using familiar analytical tools from linear algebra. In particular it is of interest to ask how different assumptions about our evolutionary process change the matrix $P$ and hence what predictions we can make about a population. An example might be to ask how does allowing mutations through neutral spaces (those for which the fitness function $f$ remains constant) affect the evolutionary dynamics? It has been established that neutral spaces might have a significant impact on how populations evolve (\cite{Schaper:2012fk}). Certainly we lose the fact that the matrix $P^*$ will be found after a finite number of iterations of the process - there exist infinite walks in our Markov Chain provided a neutral space exists.

Further, as Tan and colleagues showed (\cite{tan2011hidden}), certain paths in phenotype space can be obviated by changing landscapes.  This algebraic construction could be used to analytically design landscapes to effectively steer evolution through these high-dimensional spaces, offering the possibility of new uses for old drugs to help avoid the evolution of resistance in pathologic states such as cancer or infection.

\section*{Acknowledgment and Supplementary Information}
The authors would like to thank Arne Traulsen at the Max Planck Institute for helpful discussions.  We would like to offer the code underlying this model to any interested parties openly.  If interested, please email DN.

\bibliography{mybib}{}

\begin{thebibliography}{10}
\providecommand{\natexlab}[1]{#1}
\providecommand{\url}[1]{\texttt{#1}}
\expandafter\ifx\csname urlstyle\endcsname\relax
  \providecommand{\doi}[1]{doi: #1}\else
  \providecommand{\doi}{doi: \begingroup \urlstyle{rm}\Url}\fi

\bibitem[Moran et~al.(1962)]{moran1962statistical}
P.A.P. Moran et~al.
\newblock The statistical processes of evolutionary theory.
\newblock \emph{The statistical processes of evolutionary theory.}, 1962.

\bibitem[Weinreich et~al.(2005)Weinreich, Watson, and
  Chao]{weinreich2005perspective}
D.M. Weinreich, R.A. Watson, and L.~Chao.
\newblock Perspective: sign epistasis and genetic costraint on evolutionary
  trajectories.
\newblock \emph{Evolution}, 59\penalty0 (6):\penalty0 1165--1174, 2005.

\bibitem[Weinreich et~al.(2006)Weinreich, Delaney, Depristo, and
  Hartl]{Weinreich:2006uq}
Daniel~M Weinreich, Nigel~F Delaney, Mark~A Depristo, and Daniel~L Hartl.
\newblock Darwinian evolution can follow only very few mutational paths to
  fitter proteins.
\newblock \emph{Science}, 312\penalty0 (5770):\penalty0 111--4, Apr 2006.
\newblock \doi{10.1126/science.1123539}.

\bibitem[Tan et~al.(2011)Tan, Serene, Chao, and Gore]{tan2011hidden}
L.~Tan, S.~Serene, H.X. Chao, and J.~Gore.
\newblock Hidden randomness between fitness landscapes limits reverse
  evolution.
\newblock \emph{Physical Review Letters}, 106\penalty0 (19):\penalty0 198102,
  2011.

\bibitem[Poelwijk et~al.(2011)Poelwijk, T{\u{a}}nase-Nicola, Kiviet, and
  Tans]{poelwijk2011reciprocal}
F.J. Poelwijk, S.~T{\u{a}}nase-Nicola, D.J. Kiviet, and S.J. Tans.
\newblock Reciprocal sign epistasis is a necessary condition for multi-peaked
  fitness landscapes.
\newblock \emph{Journal of Theoretical Biology}, 272\penalty0 (1):\penalty0
  141--144, 2011.

\bibitem[Hamming(1950)]{hamming:1950}
Richard~W. Hamming.
\newblock Error detecting and error correcting codes.
\newblock \emph{Bell System Technical Journal}, 29\penalty0 (2):\penalty0
  147--160, 1950.

\bibitem[Fisher(1930)]{fisher1930}
R.~A. Fisher.
\newblock \emph{The genetical theory of natural selection.}
\newblock Clarendon Press; Oxford University Press., 1930.

\bibitem[Altenberg(1997)]{altenberg1997nk}
Lee Altenberg.
\newblock Nk fitness landscapes.
\newblock \emph{Handbook of Evolutionary Computation}, 1997.

\bibitem[Strassen(1969)]{strassen1969gaussian}
V.~Strassen.
\newblock Gaussian elimination is not optimal.
\newblock \emph{Numerische Mathematik}, 13\penalty0 (4):\penalty0 354--356,
  1969.

\bibitem[Schaper et~al.(2012)Schaper, Johnston, and Louis]{Schaper:2012fk}
Steffen Schaper, Iain~G Johnston, and Ard~A Louis.
\newblock Epistasis can lead to fragmented neutral spaces and contingency in
  evolution.
\newblock \emph{Proc Biol Sci}, 279\penalty0 (1734):\penalty0 1777--83, May
  2012.
\newblock \doi{10.1098/rspb.2011.2183}.

\end{thebibliography}
\bibliographystyle{unsrtnat}
\end{document}